%% file: WernerAurell.tex
\documentclass{revtex4}

\usepackage[dvips]{graphicx}
\usepackage{amssymb,amsfonts,amsmath}

\begin{document}

\title{A computational systems biology study of the $\lambda$-lac mutants}

\author{Maria Werner and Erik Aurell }
\affiliation{Dept. Computational Biology,KTH -- Royal Institute of Technology, AlbaNova University Center, SE-106 91~Stockholm, Sweden}

\begin{abstract}
We present a comprehensive computational study of some 900 possible
  ``$\lambda$-lac'' mutants of the lysogeny maintenance switch in
  phage $\lambda$, of which up to date 19 have been studied
  experimentally (Atsumi \& Little, PNAS {\bf 103}: 4558-4563,
  (2006)).  We clarify that these mutants realise regulatory schemes
  quite different from wild-type $\lambda$, and can therefore be
  expected to behave differently, within the conventional mechanistic
  setting in which this problem has often been framed. We
  verify that indeed, within this framework, across
  this wide selection of mutants the $\lambda$-lac mutants for the
  most part either have no stable lytic states, or should only be
  inducible with difficulty. In particular, the computational results
  contradicts the experimental finding that four $\lambda$-lac mutants
  both show stable lysogeny and are inducible. This work hence
  suggests either that the four out of 900 mutants are special, or
  that $\lambda$ lysogeny and inducibility are holistic effects
  involving other molecular players or other mechanisms, or both.  The
  approach illustrates the power and versatility of computational
  systems biology to systematically and quickly test a wide variety of
  examples and alternative hypotheses for future closer experimental
  studies.
\end{abstract}

\maketitle

The study of bacteriophage $\lambda$ has played a large role in the
development of molecular biology~\cite{echols,gottesman}, and particularly in the understanding of gene
regulation~\cite{ptashneGann,ptashne}. It is a temperate phage which
can grow lytically, or remain in a lysogenic state in the host
for many generations. While lysogeny as a phenomenon was known since
the 1920'ies, and early quantitative studies centered on other
systems, coliphage $\lambda$ became \textit{the} central model system
of lysogeny since its discovery in the early 1950'ies~\cite{lederberg,
  lwoff}. Consequently, $\lambda$ lysogeny has also been the system of
choice for mechanistic explanations of gene regulation, ranging from
systematic explanations of the data to detailed mathematical models of
the kind first presented over twenty years ago~\cite{shea}. This line
of work has been taken up several other
groups~\cite{reinitz,aurell,zhou04,saiz}, with the aims to reproduce,
in a model, known phenomena, and to shed  light on particular aspects
of such systems. Together with the lac system this approach has been
the prototype for theoretical understanding of gene regulation in
prokaryotes, generally taken to be one of the corner-stones of
quantitative systems biology~\cite{Bintu2005a,Bintu2005b,alon}.

\begin{figure}[hb]
\begin{center}
\includegraphics[width=0.6\textwidth]{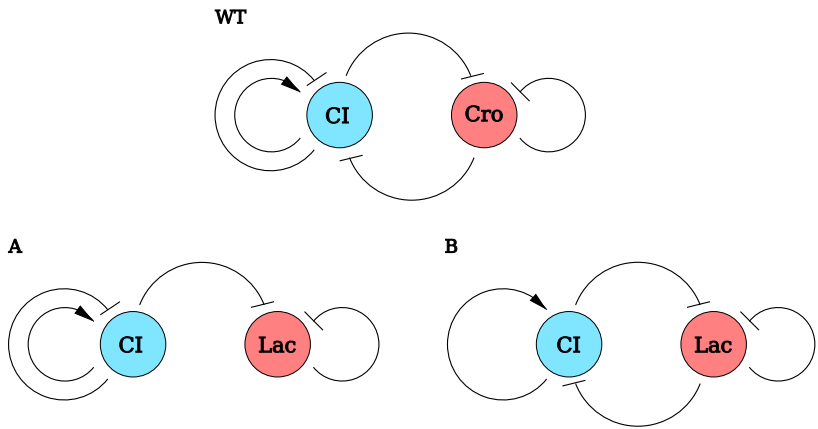}
\caption{Illustration of the three different $\lambda$ circuits.  WT) The wild-type $\lambda$ phage circuit, with auto-repression of cro and cI as well as mutual repression. CI can additionally activate its own production.  A) The circuit for mutants with an intact OR3 site, binding only CI. CI then has an intact regulatory circuit for controlling PR and PRM, compared with wild-type, while the original negative feed-back from PR, via cro, has been cut. B) The circuit for mutants with a lacO site instead of the original OR3 site. The negative feed-back from cI on PRM is then cut, while the original negative control from PR is kept.}\label{fig:scheme}
\end{center}
\end{figure}

In a recent series of experimental studies by Atsumi \& Little
\cite{atsumi,atsumi2} the lytic repressor, Cro, was replaced by the
Lac repressor, LacR.  The stated goal was to continue along the lines
of~\cite{little99} in testing the modularity of the lambda circuit,
\textit{i.e.}, in this case, to determine if stable and inducible
lysogeny is affected by a change of the lytic repressor protein. 
Therefore, the authors of \cite{atsumi,atsumi2} constructed mutants in which the
cro operon was replaced by the lac repressor operon, lacI, 
including the Shine-Dalgarno (SD) sequence for lacI. To enable
repression of PR and PL, a lac operator site, lacO, was put downstream 
of the PR and PL transcription
initiation sites. Moreover, in these new $\lambda$-lac mutants, some
carried an intact OR3 site, only binding CI and hence without any lacI
regulation of PRM. Others had the OR3 site replaced with a lacO
site, hence disrupting the cI negative control of PRM, but at the same
time allowing a negative feed-back from PR. Figure
 \ref{fig:scheme} illustrates three different circuits; the wild-type
 $\lambda$ (WT), circuit A, with intact $\lambda$ OR, and circuit B,
 where OR3 has been changed into a variant of lacO. 

 The great advantage of the new $\lambda$-lac mutants from the systems
 biology point of view is that there are potentially so many of
 them. While experimental studies have only so far been
 carried out on a fraction of all defined variants, it should be
 possible to extend the studies in~\cite{atsumi2} to many
 more. Computationally, as we will show, one can survey all variants
 in one screen, and find clear patterns.

The overall conclusion of our computational study presented here
is that the new $\lambda$-lac mutants are not explainable in standard
mathematical models of gene regulation. We believe this is of significant interest.
First, if $\lambda$ lysogeny cannot be explained, the whole program of
computational systems biology may be in trouble. We address this issue in the
Discussion. Second, at least for $\lambda$, this program is but the
formalization, in terms of defined models and equations, of what is
known or accepted in the experimental literature, often for quite some
time. Therefore, the implication would be that the functioning of the
$\lambda$ lysogeny switch is quite different to what is generally believed.

The problem can be explained by concentrating on one example, the
mutant labelled AWCF, displayed in Figure~1 in~\cite{atsumi2}. This
mutant carries a version of lacI instead of cro, a
normal $\lambda$ OR region, a strong lacO binding site at PL, an
intermediately strong lacO binding site downstream of PR, and a
(relatively) poor Shine-Dalgarno sequence. It therefore is an example 
of control circuit A in Fig. \ref{fig:scheme}. This differs from the control scheme of
wild-type $\lambda$ in that lacI does not repress cI, because lacR
does not bind anywhere in OR. One therefore expects (in the standard
framework of $\lambda$) that this mutant can be lysogenized, but that
it cannot be induced. As detailed below, the modelling conforms to
this pattern in that the kinetic equations has one equilibrium which
can be identified with lysogeny, but no equilibrium or other state at
high levels of lacR that would lead to lysis. The experimental
results (Table~1 and Fig.~3 in~\cite{atsumi2}) on the contrary
show that while indeed it can be lysogenized, it is also inducible at a UV
dose only about twice as high as in wild-type $\lambda$, releasing
almost as many phages per cell.

 \begin{figure}[ht]
\begin{center}
 \includegraphics[width=0.6\textwidth]{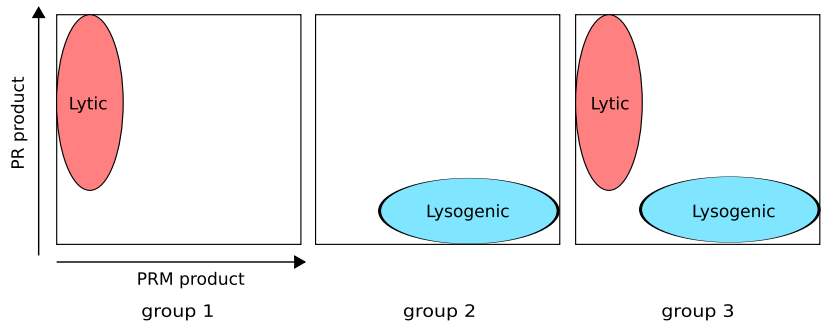}
 \caption{Three different groups into which the $\lambda$-lac mutants conceptually can be classified. In group 1, there is one equilibrium with higher level of the lytic protein than of the lysogenic protein. In group 2, there is one equilibrium with the opposite characteristics. Group 3 consists of mutants that have two equilibria, one corresponding to the lytic phase, and one with lysogenic properties. The wild-type $\lambda$ belongs to group 3, with two equilibria in the corresponding mathematical model.}\label{fig:groups}
\end{center} 
\end{figure}

 The first message of this paper, taken up again in the Discussion is
 that \textit{all} the new $\lambda$-lac mutants embody one of two
 \textit{different} schemes depending on the OR3 site.  Using
 straight-forward qualitative arguments, as above for AWCF,
 $\lambda$-lac mutants of circuit A type are expected to have stable
 lysogeny but not be easily inducible. This should also partly be true for
 circuit B, as CI then does not repress itself. Whether a stable lytic 
 equilibrium can be obtained in these mutants, should depend on the relative
 strenghts of the inserted lacO sites at PRM and PR. These conclusions are
  supported by the mathematical modelling. We categorize the mutants 
into three different groups, according to their equilibria, illustrated by 
Fig. \ref{fig:groups}. Either the mutants have only the lytic or the lysogenic equilibrium 
(group 1 or 2), or they have two equilibria and can be placed in group 3, like the wild-type phage. 

 The results of Atsumi \& Little have, we believe, implications
 far beyond what is actually claimed in~\cite{atsumi2}. In a
 well-engineered control system, a module can be replaced by another
 module of the same characteristics without changing overall system
 functionality. But one cannot, normally, replace a module with
 another of \textit{different} characteristics without either the
 system being stable against these changes, or the perturbations being
 taken care of by another additional layer of control. It is the
 first possibility that we can test in the mathematical model with
 negative results. 

\section{Results}

\subsection{Equilibria in wild-type $\lambda$}

The wild-type circuit is known to have two stable equilibria, one
lysogenic and one lytic, and the level of CI in lysogeny is fairly well
established \cite{aurell, dodd}. In the Atsumi \& Little experiments,
the PRM promoter was weakened to half its strenght,
although still enabling the lysogenic state \cite{atsumi2}. Therefore,
our wild-type lambda system was calibrated to allow a lysogenic equilibrium 
 at $\approx $ 252 CI with the original promoter strenghts, which
still allows a lysogenic equilibrium for the weaker promoter. For the latter 
case, lysogeny is naturally less stable, and has a
lower CI level.

\subsection{Equilibria in circuit A}

For circuit A, with the intact OR3 site, all variants of lacO at PR
and with any Shine-Dalgarno sequence essentially behave in the same
way. There is no negative control of PRM other than from cI itself,
while PR is inhibited by both cI and lacI. Consequently, the only
possible equilibrium is the lysogenic state, where PRM is
activated more than PR. The CI level in this state is constant at
$\approx$ 172 molecules, independent of the three
parameters (the two lacO sites and the SD sequence). The
 LacR level however depends on the
SD sequence and the lacO at PR. The most important
parameter for the LacR level is the SD sequence. The
better translation, the more LacR. Moreover, worse lacO at PR,
enables more transcription from PR, also resulting in somewhat higher
LacR levels. This is however only relevant at the best translational
efficiency. The LacR levels are however always very low, with a maximum
 of seven molecules for the best SD.

The mutants' equilibria were computed without IPTG in the system, and with a
concentration of $10^{-5}$ M IPTG. The impact of having IPTG present
is that the differences in LacR levels, caused by altering binding
affinities at PR for the best SD, completely
disappears. The promoter activities remain constant, with a PRM
activity of about 40 \% and PR completely silent. All mutants with
circuit A hence fall into group 2, with only one stable lysogenic
equilibrium.
\subsection{Equilibria in circuit B}

\begin{figure*}
 \includegraphics[width = 1\textwidth]{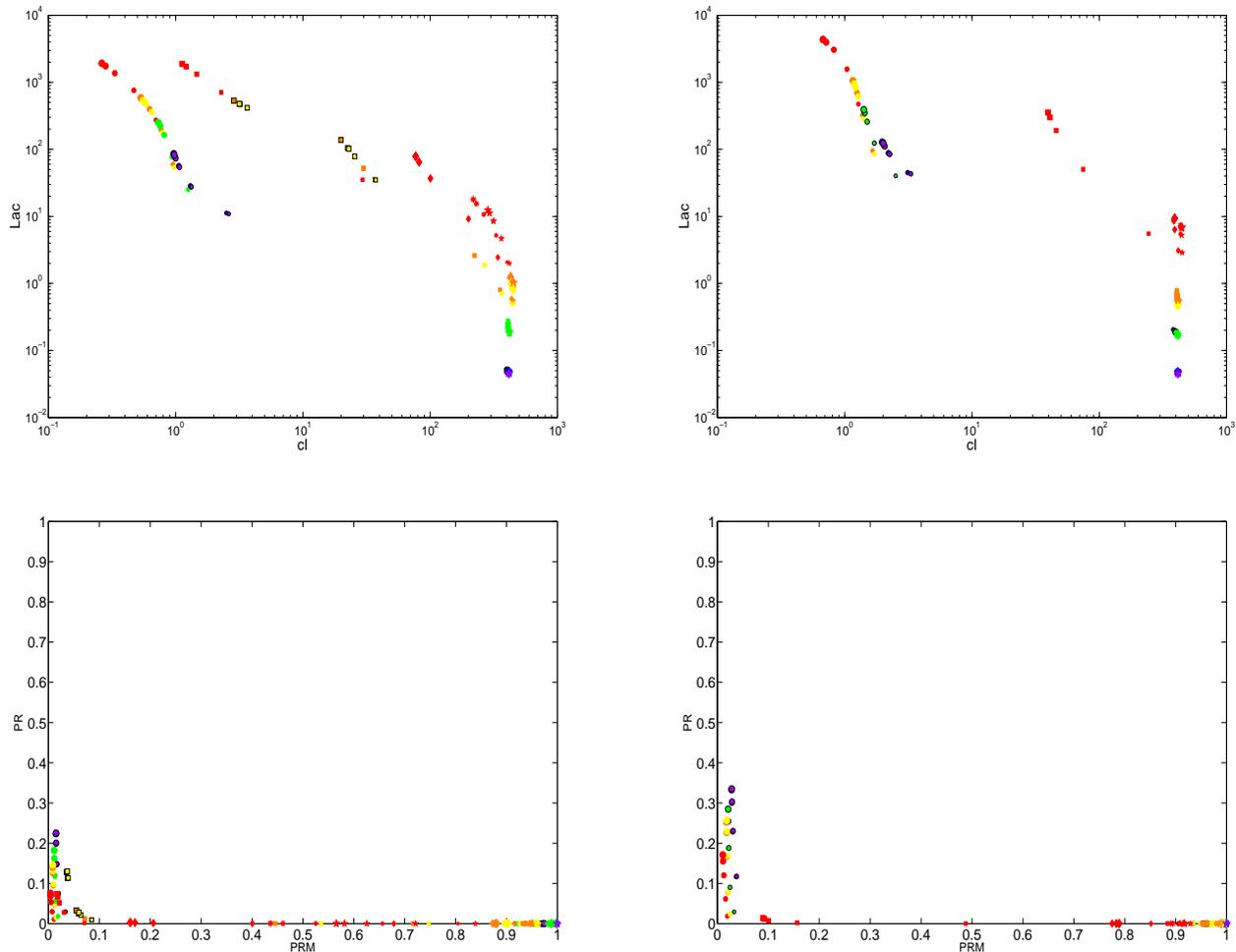}
 \caption{(AXXX) Plots showing the stable fixed points with corresponding promoter activities for the 180 mutants with the best PL lacO. The two upper plots show the fixed points at 0 IPTG (left) and $1e^{-5} M$ (right), while the lower plots show the corresponding PR and PRM activities. The mutants are coded with three parameters; colour, marker type and marker size. The lacO at PRM is indicated with marker type, going from the highest affinity $\circ$ (A) through $\Box$ (B),$\diamond$ (C),$\ast$ (D) and finally $\star$ (E). The PR lacO is indicated by the marker size, with smallest markers for the best operator (A), then increasing marker size with decreasing affinity. The Shine-Dalgarno sequence is indicated by color, with the best efficiency (A) coded as red, then decreasing with orange (B), yellow (C), green (D), blue (E) and finally purple (F). The mutants within circuit B fall into all three possible groups. There are some mutants with one lytic stable fixed point, some with both a lysogenic and a lytic fixed point, while most of them only have the lysogenic fixed point. The mutants with two fixed points are marked with black edges.}\label{fig:1XXX}
 \end{figure*}

Mutants with circuit B have the additional negative control of lacI on
 PRM, theoretically allowing a silencing of PRM and a lytic stable 
equilibrium. Depending on the combination of PRM and PR operators and the
 Shine-Dalgarno sequence, these mutants can have one or two stable
equilibria, see fig \ref{fig:1XXX}. The upper plots display the 
level of LacR and CI at the equilibria, with no IPTG (left) and with 
 $10^{-5}$ M IPTG (right), with the mutants with two equilibria marked
 with black contours. The lower plots show the corresponding PR and
 PRM activities. The mutants with this circuit can hence be in either
 one of the three groups, 1-3. 

 A stable equilibrium with higher activity of PR than PRM, appears
 when the mutant have the best PRM lacO sites (circle markers), or in some
 cases the next best, depending on the PR lacO site. The weaker the PR
 lacO is (bigger markers), the higher PR/PRM ratio. Since there is
 always a negative feedback by lacI on PR, the activity of PR is never
 above 25 \% in these 'lytic' states. The mutants with two equilibria 
are the ones with the worst SD sequence but best
 PRM lacO (XAXE or XAXF), or next best operator and better
 SD sequences (XBXC or XBXD). For the mutants with best
 PRM lacO, PR is completely silent at the lysogenic equilibrium,
 while PRM is fully active. The other equilibrium has an
 activity of PR that is higher than PRM, but with maximum 25 \%
 activity. For the mutants with next best PRM lacO (squares),
 their two equilibria are relatively close, with a promoter
 activity between 0.04-0.15 \% for both PR and PRM in both states.
Whether this state would in practice correspond to the lytic state or
 the lysogenic state for the phage is unclear.

 Most mutants however, have only one stable state, with a high PRM
 activity and low PR, corresponding to a lysogenic state in the
 wild-type $\lambda$ circuit. The activity of PR is essentially zero for
 all those cases, since a high CI level shuts down PR. The degree of
 PRM activity varies, and is in general mostly dependent on the
 SD sequence, where lower efficiency generates higher
 PRM activities. Even though the activity of PR is low, and hence the
 LacR level is low, the negative feed-back on PRM is affected from the
 occasional transcription from PR. So mutants with efficient
 SD sequences can still repress PRM in this lysogenic
 state. For many of these mutants, the PRM promoter is almost fully
 active, dramatically different from the wild-type lambda were the
 negative feed-back from cI limits the PRM activity. The CI level is 
also elevated to around 400 molecules in lysogeny, more than
 twice the lysogeny level in wild-type $\lambda$, with this weakened
 PRM promoter.

\subsection{Closer look at the 19 experimentally studied mutants}

In Little \& Atsumi's study, only 19 of the 900 mutants were isolated
and four of the mutants, all from subgroup A, were examined
more closely. The 19 isolated mutants are presented in Table 1, 
with the experimental lysogenic abilities and our computed stability results.

Two of the mutants they identified, AABF and BAAD, we predict to have
two equilibria, \textit{i.e.} place in group 3. One equilibrium has
very low activities for both promoters, and the other with full PRM
activity and a silent PR. One other mutant, ABEA, we can place into
group 1, with slightly higher PR activity than PRM. The
16 remaining mutants all have only one equilibrium, at higher PRM than
PR activity, and hence fall into group 2. For the mutants with circuit
A, the promoter activities are not dependent on IPTG levels in the
system, while for the others, the promoter activities are in general
increased with higher IPTG levels. This is to expect since the
negative feedback should be somewhat relieved when IPTG bind LacR.

\section{Discussion}
A straight-forward conclusion of this work is that the new
$\lambda$-lac mutants embody different control schemes than wild-type
$\lambda$. For this no elaborate modelling is required; case-by-case
reasoning as for the AWCF mutant in Introduction suffices. In mutants
where the $\lambda$ OR region has been left intact (of the type XWXX
where X can be any code) lacI cannot repress cI.  In mutants where OR3 has been
changed to a version of lacO (of the type XYXX where X can be any
code, and Y any code except W), cI does not autorepress itself. 
In both cases, these mutants should show exceedingly stable lysogeny
(if it can be established) because the switches should be much harder 
to throw in control schemes A and B, than in the wild-type phage.

In the quantitative modelling, every mutant is represented by a set of
two equations describing the net production of CI and LacR molecules
per unit time. An equilibrium of these equations means that the net
production of both species of molecules vanishes, such that the
concentrations are constant. A lysogenic state of the model is an
equilibrium where CI concentration is high and lacR concentration
low. The PRM promoter is then activated at a much higher rate than the
PR promoter. In models of wild-type $\lambda$ describing the
auto-activation of cI, auto-repression of cI and cro, and mutual
repression between cI and cro, there is normally both a lysogenic
equilibrium and an equilibrium where CI concentration is very low and
Cro concentration is high. In the real
system high PR activation goes together with high PL activation and
transcription of the $\lambda$ N gene, which turns on the early genes
downstream of cro and N~\cite{ptashne}, and the lysis/lysogeny
switch. The equilibrium at high Cro concentration (when it is
present in the model) therefore in reality corresponds to sustained
high level of Cro for a bacterial generation or more, sufficient to
drive the cell to lysis.

We are now ready to state the results of the modelling in more detail.
Except for a fraction of the cases in circuit B, the models of the
$\lambda$-lac mutants have, unlike wild-type $\lambda$, only one
equilibrium, and it is of the lysogenic type. This means that the
inductive process, if it is started, is self-limiting, and the switch
can never be stably thrown over to lysis.  Some of the cases in
circuit B have only a 'lytic' equilibrium, albeit with relatively low
levels of activation of PR, and these mutants should, one would guess,
therefore not be able to form lysogens.  One of these examples, mutant
ABEA, indeed did not display lysogeny \textit{in vivo}.  A few of the 
cases in circuit B have two equilibria, one
lysogenic and one 'lytic'. Of these AABF and BAAD were studied
experimentally in \cite{atsumi2}.  AABF did not show lysogeny, while BAAD showed what
Atsumi \& Little refer to as stable and unstable lysogens, where the
latter are defective in their growth properties. The four mutants which
could be lysogenized and were studied more closely all had only a
lysogenic equilibrium in the model, but could nevertheless all be
induced experimentally.

\begin{table*}[h]
\begin{center}
\caption{Table displaying the experimentally observed lysogenic properties from Little \& Atsumi's study (from Supplementary material available online \protect \cite{atsumi2}), side by side with the computed promoter activities at the equilibria. The two left columns show the experimental results at two different IPTG levels. (U = unstable, S = stable and No indicates that there were no lysogeny). The computational results are presented as PRM/PR activities, for the stable equilibria identified for each mutant.}
\begin{tabular*}{\hsize}{@{\extracolsep{\fill}}ccccccr}
\hline
 && \multicolumn{2}{c} {\textbf{Experiments}} & \multicolumn{2}{c}{\textbf{Model}}  \\
\hline
\hline
Circuit & Mutant & \multicolumn{2}{c} {Lysogeny} & \multicolumn{2}{c}{PRM/PR}  \\
\hline
& IPTG & 0 & $10^{-5}$ M &0 & $10^{-5}$ M \\
\hline
&AABF    & No     & No     & 0.02/0.08,0.96/-  &  0.04/0.12,0.99/- \\ 
&ABCD    & No     & No     & 0.94/-            &  0.99/- \\
&ABEA    & No     & No     & 0.02/0.08         &  0.09/0.01 \\
&ABEF    & No     & No     & 1/-               &  1/-  \\  
\textbf{B} &ACDA  & U   & No  & 0.17/0.002     &   0.78/- \\ 
&ACEF    & No     & No     &  1/-              &  1/- \\
&ADCA    & No     & No     &  0.53/-           &  0.88/- \\
&ADED    & S \& U & S \& U &  0.99/-           & 0.99/- \\
&AECA    & No     & No     &  0.63/-           & 0.91/- \\   
&AEDF    & No     & No     &  1/-              & 1/-  \\
\hline
&BAAD    & S \& U & S \& U &  0.02/0.02, 1/-   & 0.04/0.03, 1/- \\
&BBEF    & No     & No     &  1/-              & 1/-  \\
\textbf{B}&CBCE   & No  & No  &  1/-           & 1/-  \\
&CCDF    & No     & No     &  1/-              & 1/- \\
&DDDA    & No     & NA     & 0.70/-            & 0.98/-  \\
\hline
&AWAE    & NA     & S      & 0.41/-            & 0.41/- \\
\textbf{A}&AWCA  & S \& U & S&   0.41/-    &  0.41/-\\
&AWCD    & S      & S      &  0.41/-      &  0.41/-\\   
&AWCF    & S      & S      &  0.41/-     &  0.41/- \\  
\hline
\end{tabular*}
\end{center}

\label{tbl:mutants}
\end{table*}

We now discuss why we find these results interesting. Biological systems are
much harder to model than, say, physical system for reasons which can
be classified as \textit{(i)} unknown parameters and \textit{(ii)} 
unknown mechanisms. Models of biological systems should be robust at
least to changes in the unknown parameters, which in addition are often
likely to vary \textit{in vivo}.  Apart from its colloquial meaning,
robustness is also a technical concept in the qualitative
theory of dynamical systems, meaning that the
equilibria and other characteristics change only smoothly as
parameters are varied~\cite{arnold}.  Therefore, the behaviour of a
biological system, if described as a large system of kinetic
equations, should to a large extent be explainable in terms of its
equilibria ~\cite{kauffman,huang}. This has indeed also
been the point of view taken in previous modelling of $\lambda$, but
which does not seem to hold for the $\lambda$-lac mutants as we have
shown here.

So why is there this large discrepancy between the model and the 
experiments, and what does this imply for $\lambda$?  The first and 
maybe most obvious explanation is that the 19 mutants studied experimentally
could be a biased sample, and that if all were screened, then they
would conform more closely to the theoretical prediction. This is
perhaps unlikely, but supports the utility of the
$\lambda$-lac mutants as a test bed, and makes the case that an
experimental investigation of many variants would be most interesting.

\begin{center}
  \begin{figure}
\includegraphics[width = 0.7\textwidth]{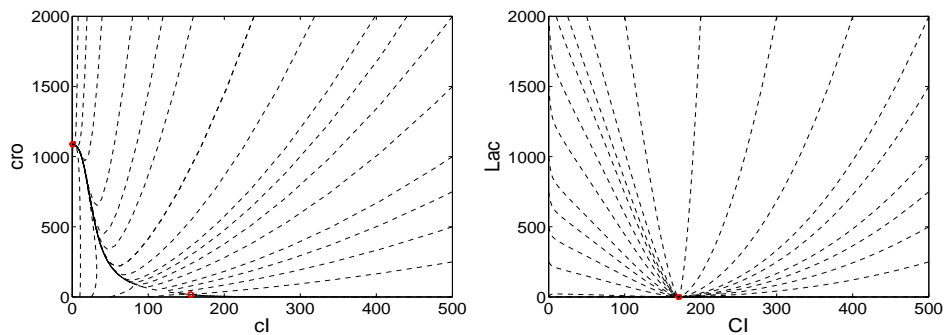}
 \caption{Phase plots for wild-type $\lambda$ phage in  \cite{atsumi2} and the AWCF $\lambda$-lac mutant. The left plot shows the phase plot for the wild-type phage, with two quilibria, one lysogenic and one lytic. The right plot shows the corresponding phase plot for the AWCF mutant, belonging to circuit A. This mutant exhibits only one equilibrium, at the lysogenic point.}\label{fig:wt}
 \end{figure}
\end{center}

A second hypothesis is that, in fact, no ``lytic'' equilibrium is necessary, 
but that instead a transient rise in Cro is enough to excite genes
down-stream of Cro and throw the switch to lysis. This contradicts
the original picture of $\lambda$ lysogeny presented in 
\textit{e.g.}~\cite{ptashne}, 
but is more in line with recent results indicating that Cro is not 
necessary for lytic induction \cite{svenningsen,dodd,dodd2001}. 
However, our comprehensive simulations of the transients 
`(``phase-space plots''), indicate that 
very few of the mutants have high Lac transients. Especially, the AWCF mutant 
that experimentally behaves similarly to the wild-type, 
indeed have no transient rise in LacR (see Fig. \ref{fig:wt}).  

A third hypothesis is that some basic assumption of the modelling may
be incorrect. Is this concievable? Perhaps so. One observation is that
the time scales of LacR binding/unbinding have recently been measured
in single-cell experiments, and are as long as on the order of a
minute~\cite{elf2007}. The implied time scales for a looped complex
between OR and OL to disassociate may therefore be on the order of a
bacterial generation, in which case \textit{in vivo} occupancy of the
operator sites by CI would be far from in equilibrium with the
instantaneous CI concentration.  Lysogeny would then not be a
quasi-equilibrium state, but an intermittently controlled process,
where periods of dilution and degradation alternate with sharp bursts
of protein synthesis, and such a system would be more easily
inducible.  Other mechanisms are certainly also possible. The lesson
of this study is that the module replacement approach of~\cite{atsumi2}
shows that the $\lambda$ lysogeny switch may in fact not be a well understood
phenomenon.

 \section{Materials and Methods}

As already outlined, in the $\lambda$-lac
mutants in \cite{atsumi2}, the lacI gene of the \textit{E. coli} lactose system has been substituted for the cro gene of $\lambda$. LacO sites were introduced to
 maintain transcriptional control, and SD sequences for the lac repressor were inserted. Altogether, Atsumi \& Little used 
five variants of lacO, labelled (A-E), and six variants
of SD, labelled (A-F). In total, the two 
circuit constructs resulted in 900 possible mutants, identified by a four-letter code
as in the example $AWCF$ in the introduction. 

Since the experimental genetic constructs fall into two categories,
(circuit A and B in Fig. 1), and the wild-type lambda
consists of yet another circuit, three separate models were
constructed. They differ since the genetic architecture and regulatory
mechanisms differ while they are all thermodynamic models, in line
with the earlier studies of the lambda switch \cite{shea,reinitz,aurellPRL}. 
A short description of the general setup and model assumptions are presented below. 
For a complete description of all model parameters, see Appendix B and C.

\subsection{General setup} 

The models all include the left and right
operator sites, OL and OR, both binding CI and Cro/LacR. The models allow looping
formations between OL and OR, caused by CI forming tetramer and
octamer structures \cite{dodd2001}. The equilibria are obtained from 
solving the equations describing the net changes in CI and LacR levels, 
see Eq 1 and 2. CI is a stable protein and is not degraded, only diluted 
with a continous rate corresponding to a generation time of $\tau_{div}$. LacR is 
besides diluted by cell division, also degraded with a half-life of $\tau_{deg}$. 
Production of CI from PRM can take place with three different rates, $R_{PRM}^u$ for 
the unactivated promoter, $R_{PRM}^{nl}$ for an activated PRM in a non-looped configuration 
and finally $R_{PRM}^{l}$ for the activated promoter in a looped configuration. The PR 
promoter has one constant production rate of $R_{PR}$.

\begin{table*}[hb]
\begin{align}
\frac{dCI}{dt} &= -CI \frac{ln 2}{\tau_{div}} +  \frac{1}{2} \times S_{CI} (R_{PRM}^u P_{PRM}^u + R_{PRM}^{nl} P_{PRM}^{nl} + R_{PRM}^l P_{PRM}^l)\\
\frac{dLacR}{dt} &= -LacR( \frac{ln 2}{\tau_{div}} + \frac{ln 2}{\tau_{lac}})+  S_{LacX}  R_{PR} P_{PR}
\end{align}
\end{table*}

Unactivated transcription from PRM occurs when OR3 and OR2 
are unoccupied, while activated transcription requires a CI dimer to 
be bound at OR2. PR is active with a constant rate as long as OR1 and
 the lacO at PR remains unoccupied. 

\subsection{Looping configurations}

In earlier models of the $\lambda$ phage, the fully occupied OL 
or/and OR state was treated as one state, with the assumption that 
OR1 and OR2 forms a cooperative binding \cite{aurell,shea,saiz,dodd}. 

However, theoretically there are cooperative bindings either
between CI bound at OR1 and OR2 or at OR3 and OR2, meaning that in
principle two separate configurations exist for the fully occupied
state. These configurations become important to separate when looping is 
modeled (for details see Appendix A). Our state description 
includes both these configurations. 

\include{Appendix}

\end{document}

%% file: Appendix.tex
\appendix
\section{Looping configurations}

CI molecules form dimers that can bind into each site at the left and right 
operators, OL and OR. If CI dimers occupy two neighbouring sites, they can 
interact and bind cooperatively. Dimers bound at OL and OR can also interact 
with each other, if the DNA forms a loop and allows the molecules to come 
close enough to each other \cite{dodd2001,dodd}. If there exist a cooperative 
interaction between two neighbouring dimers at OR and OL, these can together 
form an octamer structure, while single dimers at OL and OR can together form 
a tetrameric structure. 

The promoter activities of PRM and PR depend on the CI 
(and LacR) binding patterns at OR and OL, and each possible configuration 
of bound CI and LacR needs to be evaluated. For example, when OR is fully 
occupied by CI, \textit{i.e} CI dimers are bound to all three sites, 
there are in fact two different configurations, depending on if CI dimers 
at OR1 and OR2 are bound cooperatively, or dimers at OR2 and OR3. In earlier 
models of the $\lambda$ phage switch, before looping between OL and OR had been
identified, these two configurations do not need to be separated in the description.
The have the same statistical weight, and the same consequence regarding promoter 
activities \cite{shea,reinitz}. 

\begin{center}
  \begin{figure}[hb]
\includegraphics[width = 0.5\textwidth]{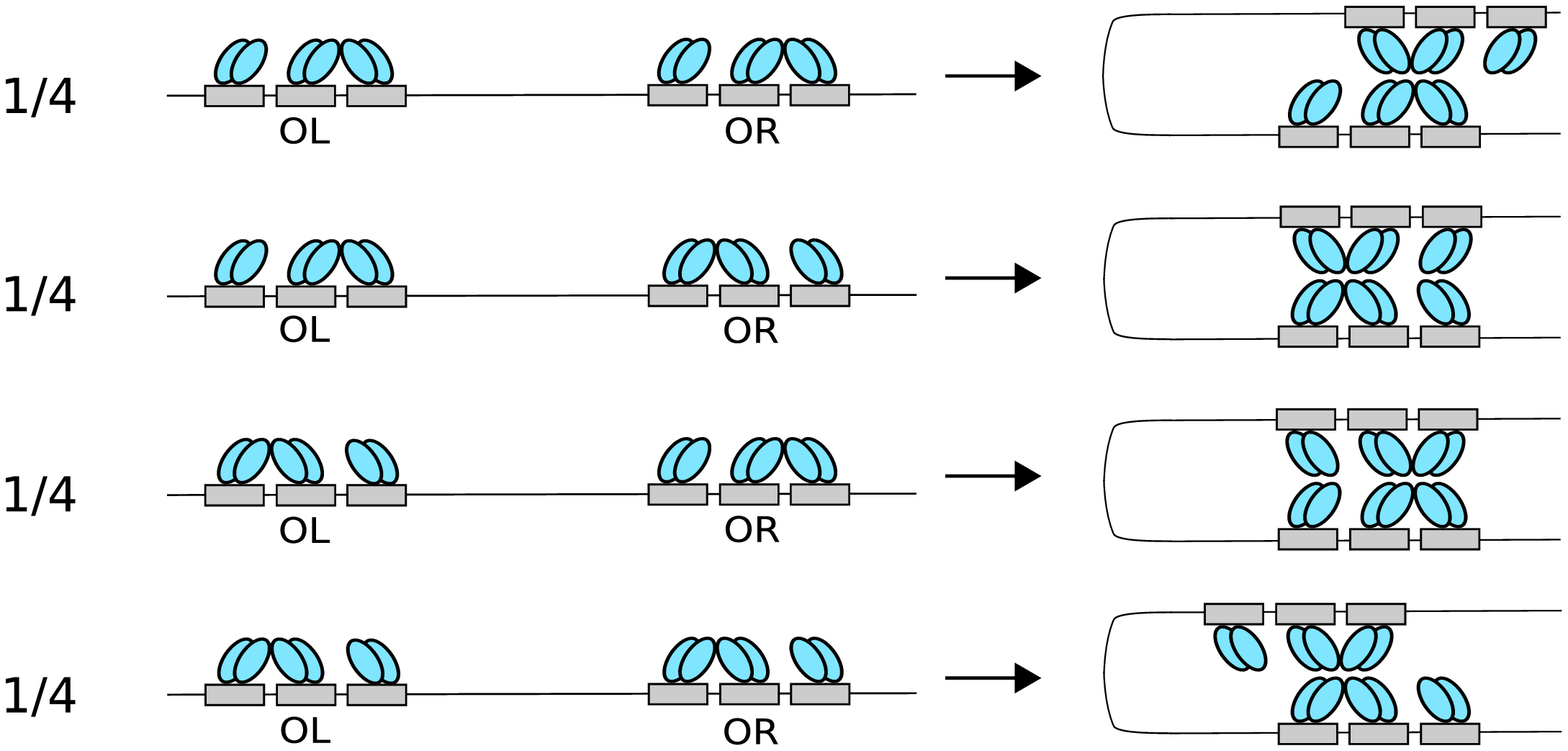}
 \caption{Illustration of the four possible configurations when Ol and OR are both fully occupied by CI. They are all equally probable, \textit{i.e.} occur with weight 1/4.}\label{fig:conf}
 \end{figure}
\end{center}

However, in a model that acknowledges looping, the two possible fully occupied 
configurations need to be treated separately, since they allow different looped 
structures to form. Fig. \ref{fig:conf} illustrates the different possible 
configurations, in the non-looped and the looped state. If CI at OR1 and OR2 
are bound cooperatively, and OL2-OL3 likewise, only the octameric structure should
be able to form. CI bound at OR3 and OL1 can then never interact and form a tetramer 
since they are placed on opposite sites of the octamer. The same occurs if the opposite scenario, with OL1-OL2 and  OR3-OR2 bound cooperatively. So only two
out of the four configurations of fully occupied operators actually allow both an octamer and a tetramer to be formed. 

When the cooperative binding energies between neighbouring bound CI are 
set equal for all interactions, the four configurations are equally 
probable (each with weight 1/4). In principle, this way of counting 
configurations, according to the ``tilting'' of the CI dimer, also means 
that all configurations could be separated into
different tilted versions. However, it is only in the fully occupied
state that it becomes necessary to keep them separated. For the configurations 
where two CI dimers are bound as neighbors but tilted away from each other, 
\textit{i.e.} three out of four scenarios, they are so much less probable 
than the configuration with cooperative binding, that they can be neglected.

\section{Parameters}

\subsection*{Lac repressor}

The wild-type LacR forms relatively stable tetramers \cite{wilson}, 
while the dimeric form of LacR used in Atsumi \& Little's study , is 
most likely less stable. The half-life is here taken to be 20 min \cite{platt}, and the repressor dimerization constant is set to 80 nM \cite{atsumi2}.

The dimeric mutant is most likely similar in properties to the -32 aa dimeric
mutant presented in Chen and Matthews study from 1992
\cite{chenJ}. They measured the apparent operator binding dissociation
constant for this mutant to be lower, but concludes that it is not the
DNA binding properties that are altered, but the dimerization constant
\cite{chenJ}. Therefore, it is assumed to bind as the wild-type repressor 
to the optimal lacO site, with a Kd of $1.2*10^{-11}$ M \cite{chenJ}. 
The affinity for the Lac repressor to the various mutated operator 
sites was obtained from a study by Betz \textit{et al.} in 1986 \cite{betz}.  The dissociation constants for the three first mutations (B,C,D) were experimentally determined in \cite{betz}. For the fourth sequence, the dissociation constant were not measure directly, only the $\beta$-galactosidase activity, \textit{i.e.} here the measure of repression of promoter due to binding at the mutated lacO site. By correlating the 15 total measured dissociation constants with their corresponding $\beta$-galactosidase activites, also the fourth lacO site affinity was estimated. The binding strengths of the four mutated lacO sites, relative the best operator A, are [ 4.17 10.83 16.67 19.72] (B-E).

Regarding IPTG interactions with LacR, one IPTG can bind each 
LacR monomer, either the free monomer, or each monomer in a multimer. 
IPTG bound to the repressor lowers its
affinity for lacO sites, while nonspecific binding is not affected \cite{bourgeois}. 
 The operator affinity for IPTG-bound LacR was here estimated from earlier 
theoretical models of LacR-IPTG interactions, with $\approx$ 5 and 30-fold 
reduced operator affinities (one or two IPTG)\cite{ogorman}. 

\begin{center}
  \begin{figure}[hb]
\includegraphics[width = 0.5\textwidth]{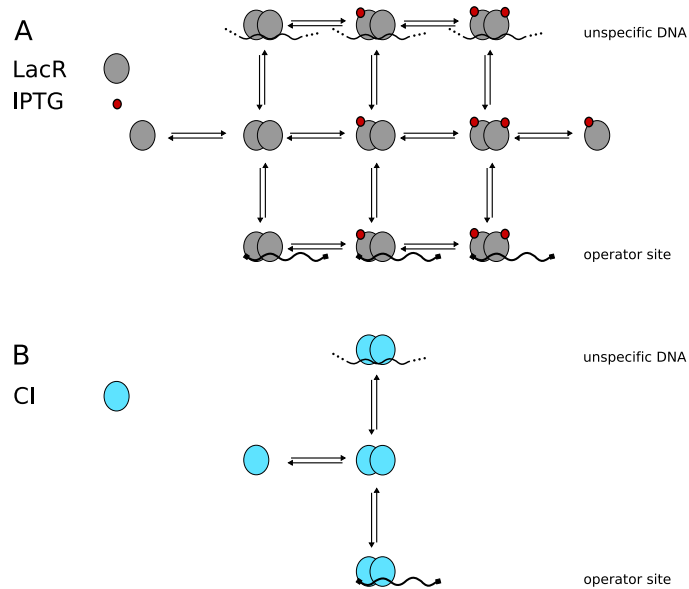}
 \caption{Illustration of all interactions between molecular species in the model. LacR dimerizes and each lacR unit can bind IPTG. LacR dimers with, and without, IPTG can bind to specific operator sites or non-specifically to DNA. CI also dimerizes and can bind to operator sites or to DNA non-specifically.}\label{fig:lacr}
 \end{figure}
\end{center}

The IPTG affinity for the LacR monomer is changed when the repressor is 
bound at an operator site. Moreover, Operator-bound LacR experience a positive cooperative binding, although the IPTG affinity for repressor is being
reduced about 20-fold \cite{ogorman,dunaway}. Our model assumes that
the dimerization of Lac repressor is independent of the IPTG binding. Fig. \ref{fig:lacr} shows all interactions between LacR and IPTG, operators and DNA, that are included in the model. 

\subsection*{CI and Cro}

The CI repressor dimerizes with a dissociation constant, $K_{dCI}$, of $2 \times 10^{-8}$M \cite{dodd,reinitz}, and binds DNA as a dimer. The CI repressor affinities for the three OR sites have been measured by Koblan \& Ackers and for OL by Senear
\textit{et al.}. In our model we have used the free energy
measurements for the OR, taken at 37 degrees
\cite{koblan}.  The affinities at OL sites were estimated from 
measurements at 20 $^{\circ}$C with corrections for higher temperature following the OR measurements \cite{koblan,senear}. The free energies for forming an octameric or tetrameric looped structure were taken from \cite{dodd}.

In the wild-type $\lambda$ model the Cro binding affinities were taken from \cite{aurellPRL,takeda89}. The unspecific binding of Cro has been taken as in \cite{aurellPRL}. Cro dimerizes strongly, with a dissociation constant of $1.2 \times 10^{-5}$ \cite{aurellPRL} and the life-time of Cro is set to be 43 minutes \cite{pakula}. The number of Cro molecules produced from each cro mRNA has been estimated to 20, follwing analysis in \cite{aurell}.

\subsection*{Transcriptional rates}
In the wild-type circuit, the unspecific binding of CI was used as
free parameter, to calibrate the PR and PRM rates in order to set a
lysogenic fix-point of $\approx$ 252 CI and 3 Cro
\cite{aurell,dodd}. The relative rates for PRM, in its three different
states, and PR, was taken from experimental measurements
\cite{dodd}. These calibrated rates were then used in the $\lambda$-lac models for valid comparisons. 
 
\subsection*{Translational efficiencies}

The translational efficiency $S_{CI}$
is assumed to be one CI molecule per transcript \cite{aurell,aurellPRL}. The translational efficiency of the lacI, $S_{LacX}$, is dependent on the Shine-Dalgarno sequence, where the sequences were said to allow gradually lower
translational efficiency, from sequence A to F \cite{atsumi2, gardner}.
The best SD sequence is set to generate 23 repressors from each mRNA \cite{kennell}. 

 To estimate the actual relative translational efficiencies
we applied the efficiency-matrix established from experimental data by
Barrick \textit{et. al} \cite{barrick}. The six sequences were
accordingly ranged from higher to lower efficiencies. The obtained number of translations per mRNA, for the optimal SD and the five mutated sequences, is then [23 3.77 3.31 1.24 0.35 0.33] (A-F).

\section{Tables}

\begin{table}[Ht]
\begin{center}\caption{All parameters included in the model of the $\lambda$-Lac phages, including notation used in main article, plus references. (Ref. 1 and 2 are theoretical studies with further references to be found within.)}
\begin{tabular*}{\hsize}{@{\extracolsep{\fill}}llll}
\\
\hline
&&&\\
\large\textbf{Parameter}& \large\textbf{Notation} & \large\textbf{Value} & \large\textbf{Reference} \\
&&&\\
\hline
\hline
cell volume & V & $2 \times 10^{-15} L  $ & \cite{aurellPRL,aurell}\\
length of bac DNA & $L_{DNA}$ & $5*10^{6} $bp & \cite{aurell}\\
lifetime CI & $\tau_{CI}$ & $ \inf $  & \cite{reinitz}\\
lifetime LacR & $\tau_{LacR}$ & 20 min  & \cite{platt}\\
cell generation time & $\tau_{div}$ & 34 min & \cite{aurellPRL,aurell}\\
temperature & $k_BT$ & 0.617 kcal/mol & \cite{reinitz,aurellPRL,aurell}\\
\hline
 \multicolumn{2}{c}{\textbf{Translational efficiencies}} & \textbf{[no]}&\\
\hline
translation cI & $S_{CI}$ & 1 & \cite{aurellPRL,aurell} \\
translation lac (A)& $S_{LacA}$ & 23 & \cite{kennell}\\
translation lac (B)& $S_{LacB}$ & 3.77 & \cite{barrick}\\
translation lac (C)& $S_{LacC}$ & 3.31 & \cite{barrick}\\
translation lac (D)& $S_{LacD}$ & 1.24 & \cite{barrick}\\
translation lac (E)& $S_{LacE}$ & 0.35 & \cite{barrick}\\
translation lac (F)& $S_{LacF}$ & 0.33 & \cite{barrick}\\
\hline
 \multicolumn{2}{c}{\textbf{Transcription rates}} & \textbf{[1/s]}&\\
\hline
$PR$ rate & $R_{PR}$ & 1.0171  & fit to \cite{dodd}\\
$PRM$ rate, unactive& $R_{PRM}^u$ &  0.0433 &  fit to \cite{dodd}\\
$PRM$ rate, unlooped & $R_{PRM}^{nl}$ &  0.3468 &   fit to \cite{dodd}\\
$PRM$ rate, looped & $R_{PRM}^l$ &  0.2552 &  fit to \cite{dodd}\\
\hline
 \multicolumn{2}{c}{\textbf{Dimerization and inducer binding}} & \textbf{[M]}&\\
\hline
Dimerization CI & $K_{dCI}$& $1.5*10^{-8} $ &\cite{aurellPRL, dodd}\\
Dimerization  LacR & $K_{dLac}$ & $8* 10^{-8}$ & \cite{atsumi2}\\
LacR binding IPTG & $K_{RI}$ & $1.6*10^{-6} $& \cite{chenJ}\\
$LacR_2-O$  binding IPTG& $K_{IdnaR}$& $3.2*10^{-5}$ & \cite{ogorman,dunaway}\\
$LacR_2I-O$ binding IPTG&$K_{IdnaRI}$ & $1.6*10^{-5}$ & \cite{ogorman,dunaway}\\
\hline
 \multicolumn{2}{c}{\textbf{LacR DNA binding}} & \textbf{[M]}&\\
\hline
$LacR_2$ unspec binding& $dg_{Lacun}$ & $1*10^{-4}$ &\cite{ogorman}\\
$LacR_2$ spec binding (A) & $K_{dnaRA}$ & $1.2*10^{-11} $ & \cite{chenJ} \\
$LacR_2$ spec binding (B) & $K_{dnaRB}$ & $5*10^{-11} $ & \cite{betz} \\
$LacR_2$ spec binding (C) & $K_{dnaRC}$ & $1.3*10^{-10} $ & \cite{betz} \\
$LacR_2$ spec binding (D) & $K_{dnaRD}$ & $2*10^{-10} $ & \cite{betz} \\
$LacR_2$ spec binding (E) & $K_{dnaRE}$ & $2.4*10^{-10} $ & \cite{betz} \\
$LacR_2I$ spec binding & $K_{dnaRI}$ & $5.6 \times K_{dnaRX}$&  \cite{ogorman} \\
$LacR_2I_2$ spec binding & $K_{dnaRI2}$ & $30.5 \times K_{dnaRX}$ &  \cite{ogorman} \\
\hline
 \multicolumn{2}{c}{\textbf{CI DNA binding}} & \textbf{[kcal/mol]}&\\
\hline
$CI_2$ unspec binding & $dg_{CIun}$ & -3.3 &  free parameter \\
$\Delta G_{or1}$ & $dg_{or1}$ & -12.5 & \cite{koblan}\\
$\Delta G_{or2}$ & $dg_{or2}$ & -10.5 & \cite{koblan}\\
$\Delta G_{or3}$ & $dg_{or3}$ & -9.5 & \cite{koblan}\\
$\Delta G_{ol1}$ & $dg_{ol1}$ &  -13.1 & \cite{senear,koblan} \\
$\Delta G_{ol2}$ & $dg_{ol2}$ &  -11.9 & \cite{senear,koblan}\\
$\Delta G_{ol3}$ & $dg_{ol3}$ &  -11.5 & \cite{senear,koblan}\\
$\Delta G_{olcoop}$ & $dg_{coop} $ &-2.8 &  \cite{koblan} \\ 
$\Delta G_{oct}$ & $dg_{oct}$ & -0.5 & \cite{dodd}\\
$\Delta G_{tet}$ & $dg_{tet}$ & -3  & \cite{dodd}\\
\hline
\end{tabular*}
\end{center}
\end{table}

\begin{table}
\begin{center}\caption{Additional parameters included in the model of the wild-type phage, including notation used in main article and references. The general parameters and LacR parameters are the same as in Table 1 supplementary materials.} \label{table_label}
\begin{tabular*}{\hsize}{@{\extracolsep{\fill}}llll}
\\
\hline
&&&\\
\large\textbf{Parameter}& \large\textbf{Notation} & \large\textbf{Value} & \large\textbf{Reference} \\
&&&\\
\hline
\hline
lifetime Cro & $\tau_{lac}$ & 43 min  & \cite{pakula}\\
translation cro& $S_{cro}$ & 20 [no] & \cite{aurellPRL}\\
Dimerization cro & $K_{dcro}$& $1.2*10^{-5} [M]$ &\cite{aurellPRL}\\
\hline
 \multicolumn{2}{c}{\textbf{cro DNA binding}} & \textbf{[kcal/mol]}&\\
\hline
 $Cro_2$ unspec binding  & $d_{gCroun}$& -6.5 & \cite{aurellPRL}\\
 $\Delta G_{or1}$ & $dg_{or1}$ & -14.4 & \cite{aurellPRL}\\
$\Delta G_{or2}$ & $dg_{or2}$ & -13.1 & \cite{aurellPRL}\\
$\Delta G_{or3}$ & $dg_{or3}$ & -15.5& \cite{aurellPRL}\\
$\Delta G_{ol1}$ & $dg_{ol1}$ &  -14.5 & \cite{aurellPRL,takeda89} \\
$\Delta G_{ol2}$ & $dg_{ol2}$ &  -13.9 & \cite{aurellPRL,takeda89}\\
$\Delta G_{ol3}$ & $dg_{ol3}$ &  -13.4& \cite{aurellPRL,takeda89}\\
\hline
\end{tabular*}
\end{center}
\end{table}